\documentclass[11pt]{article}
\usepackage{bm,amsmath,amssymb,color,centernot,multicol,fancybox,mathrsfs,graphicx,here,chicago,neil}

\begin{document}

\allowdisplaybreaks
\baselineskip=20pt

\title{Stochastic volatility model with range-based correction and leverage}

\author{Yuta Kurose\footnote{Faculty of Engineering, Information and Systems, University of Tsukuba, 1-1-1 Tennodai, Tsukuba-shi, Ibaraki 305-8573, JAPAN. E-mail: kurose@sk.tsukuba.ac.jp}}

\date{October, 2021}

\maketitle

\thispagestyle{empty}

\begin{center}
{\bf Abstract}
\end{center}

\noindent
This study presents contemporaneous modeling of asset return and price range within the framework of stochastic volatility with leverage. 
A new representation of the probability density function for the price range is provided, and its accurate sampling algorithm is developed. 
A Bayesian estimation using Markov chain Monte Carlo (MCMC) method is provided for the model parameters and unobserved variables. 
MCMC samples can be generated rigorously, despite the estimation procedure requiring sampling from a density function with the sum of an infinite series. 
The empirical results obtained using data from the U.S. market indices are consistent with the stylized facts in the financial market, such as the existence of the leverage effect. 
In addition, to explore the model's predictive ability, a model comparison based on the volatility forecast performance is conducted.

\vspace{\baselineskip}


\newpage
\setcounter{page}{1}
\section{Introduction}
Estimating the variance or volatility of an asset return observed in the financial market has been an attractive research topic for many authors. 
Stochastic volatility (SV) model and generalized autoregressive conditional heteroskedasticity (GARCH) model are two major statistical models for asset return through describing the dynamics of the variance using state space representation (\cite{Bauwens_etal(2012)}). 

Many scholars have studied a model-free approach for estimating the volatility of asset return using price range since the 1980s. 
Price range is defined as the logarithmic difference between the highest and lowest asset prices of daily transactions, and \cite{Feller(1951)} derived the distribution as the sum of an infinite series.
\cite{Parkinson(1980)} derived the $p$-th moment of the range and provided an estimator of the variance, although \cite{GarmanKlass(1980)} pointed out that the ranges observed in the financial market may have (downward) bias due to price discretization noise and non-trading hours. 
\cite{GarmanKlass(1980)}, \cite{RogersSatchell(1991)}, \cite{Kunitomo(1992)}, and \cite{YangZhang(2000)}
explored such bias and proposed volatility estimators based on a range (also see \cite{Molnar(2012)}). 
On the other hand, some authors took a model-based approach for volatility estimation using the price range. 
For example, \cite{Gallant_etal(1999)}, \cite{Alizadeh_etal(2002)}, and \cite{BrandtJones(2005)} investigated SV models using a daily range of the log asset process. 
\cite{Horst_etal(2011)} presented a joint model for daily asset return, the highest and lowest log prices within the context of SV models. 
\cite{Chou(2005)}, \cite{BrandtJones(2006)}, and \cite{Chen_etal(2008)} presented GARCH-related models that make use of the range. 
I notice that these articles approximate the likelihood function as an easy density or sum of a finite series since the distribution function of the range is complicated, as described before, and exact computation of the likelihood is difficult. 
To the best of my knowledge, the rigorous evaluation of the likelihood is not found in the past papers.  

Due to the availability of high-frequency data in the financial market, the realized measure has been provided by many authors as the proxy for a latent variable of interest using such data in recent decades. 
\cite{AndersenBollerslev(1998)} proposed a daily volatility estimator known as ``realized volatility,'' the sum of the finitely sampled squared returns over a day. 
Many pieces of literature showed that it is a very informative measure for volatility. 
As pointed out by many authors (\cite{Bauwens_etal(2012)}), however, it is known that high-frequency data is contaminated by noise, and proxies for unobserved components using such data are biased. 
For example, \cite{MartensDijk(2007)} and \cite{ChristensenPodolskij(2007)} discussed the bias of high-low range for the short interval of a day. 
As a result, many studies addressed this topic and presented various realized measures for accurately estimating unobserved variables of interest. 
\cite{MartensDijk(2007)}, \cite{ChristensenPodolskij(2007)}, and \cite{Christensen_etal(2009)} proposed ``realized range'' volatility estimators based on short interval ranges. 
Other than this model-free approach, some authors took a model-based approach for estimating the variance of daily asset return using the return and the corresponding realized measure simultaneously. 
\cite{Takahashi_etal(2009)} and \cite{DobrevSzerszen(2010)} extended the SV model framework by adding an observation equation that describes the linkage between realized volatility and unobserved volatility. 
It is referred as realized stochastic volatility (RSV) model, and related models were studied by \cite{KoopmanScharth(2013)}, \cite{Shirota_etal(2014)}, \cite{ZhengSong(2014)}, and \cite{Takahashi_etal(2016)}. 
Realized GARCH model was introduced as an extension of the GARCH model to relate latent volatility and the corresponding realized measure, similarly as RSV models (e.g., \cite{Hansen_etal(2012)}, \cite{Watanabe(2012)}, \cite{Wang_etal(2019)}, and \cite{Chen_etal(2021)}). 

Although approaches based on realized measures successfully and accurately estimate the latent volatility of an asset return, 
\cite{DegiannakisLivada(2013)} noted that realized volatility is a more accurate estimator than range-based one only when the intraday dataset is available at high sampling frequency. 
That is, the dependability of these approaches is determined by the transaction frequency of the asset. 
On the other hand, even if the asset is not liquid and has low transaction frequency, the financial market always has the open, highest, lowest, and closing prices of the asset. 
Thus, the range-based approach for the estimation of latent volatility is appealing in light of data availability. 
Additionally, collecting and storing the few transaction prices is a very light burden compared with acquiring and storing the detailed intraday transaction information and constructing the corresponding realized measure from it. 

As stated above, there are two major problems about the usage of the range for volatility estimation in the previous studies: 
(1) leverage effect and (2) calculation of the likelihood function. 
In this study, I tackle these problems and propose a new framework that models the return and the price range jointly in the context of the SV model with leverage. 

The leverage effect is a phenomenon in the stock market: a negative relationship between the return of an asset at date $t$ and the variance of the return at day $t+1$. 
The Bayesian estimation method of the SV model with leverage was discussed by \cite{Jacquier_etal(2004)} and \cite{Omori_etal(2007)}. 
As far as the author knows, past research did not deal with SV model using range data that incorporates the leverage effect. 

In this paper, I express the range's density function as the sum of an infinite series other than the one derived by \cite{Feller(1951)}. 
The two representations enable us to generate samples from the density rigorously without computing the sum of the infinite series.
Using a Bayesian approach, I apply this sampling method to the proposed model and implement a generation algorithm for latent volatilities via the Markov chain Monte Carlo (MCMC) method. 

The rest of this article is organized as follows. 
In Section 2, I investigate the density of the range and sampling from the density. 
Bias correction of the range is discussed, and the SV model with range-based correction and leverage is proposed. 
Section 3 introduces a Bayesian estimation method for the model that make use of an efficient MCMC algorithm. 
Section 4 contains an empirical analysis that uses financial return and price range data from the U.S. stock indices.
The findings of the paper are presented with a conclusion in Section 5.

\section{High-low range and volatility}
\subsection{Distribution of the range}
I consider a continuous time process, 
\begin{equation}
dp(s) =\sigma(s)dW(s),
\end{equation}
where $p(s)$ is the logarithm of the price of an asset observed in the financial market at time $s$, 
$\sigma^2(s)$ denotes the instantaneous volatility, 
and $W(s)$ is a standard Wiener process. 
I also assume that $\sigma^2(s)$ is statistically independent of $W(s)$. 

The main objective of this study is to estimate the true volatility or integrated volatility for a day $t$ defined as 
\begin{equation}
\sigma^2_t =\int_{t}^{t+1} \sigma^2(s) ds.
\end{equation}
Let the highest price of the day $t$ be $H_t$, 
and the lowest price of the day $t$ be $L_t$. 
I define the price range in day $t$, $r_t$, as 
the log difference between the highest and the lowest value, i.e., $\log H_t -\log L_t$. 

\cite{Feller(1951)} showed that the density function of range $r_t$ given $\sigma^2_t$ is 
\begin{equation} \label{eq:2.1-a}
f_{\mathrm{range}}(r_t|\sigma^2_t) =8\sum_{n_d=1}^\infty (-1)^{n_d-1} \frac{n_d^2}{\sqrt{2\pi\sigma_t^2}}\exp\bigg( -\frac{n_d^2r_t^2}{2\sigma_t^2} \bigg),
\end{equation}
and that 
\begin{equation}
f_{\mathrm{range}}(r_t|\sigma^2_t) =\bigg( \frac{2}{\pi} \bigg)^{\frac{1}{2}} r_t^{-1} L'\bigg( \frac{r_t}{2\sigma_t} \bigg), 
\end{equation}
where $L(x) =(2\pi)^{\frac{1}{2}} x^{-1} \sum_{n_d=1}^\infty \exp\big\{ -\frac{(2n_d-1)^2\pi^2}{8x^2} \big\}$ is the Kolmogorov-Smirnov distribution function. 

\cite{Feller(1948)} stated that $L(x)$ can be written in two equivalent forms as 
\begin{align}
L(x) &=(2\pi)^{\frac{1}{2}} x^{-1} \sum_{n_d=1}^\infty \exp\bigg\{ -\frac{(2n_d-1)^2\pi^2}{8x^2} \bigg\} \\
&= 1- 2\sum_{n_d=1}^\infty (-1)^{n_d-1} \exp( -2n_d^2x^2 ). \label{eq:2.1-b} 
\end{align}

It is straightforward that the density function (\ref{eq:2.1-a}) corresponds to the distribution function represented in equation (\ref{eq:2.1-b}). 
I also obtain another representation of the density as follows: 
\begin{equation}
f_{\mathrm{range}}(r_t|\sigma^2_t) =8\sum_{n_d=1}^\infty \bigg\{ \frac{(2n_d-1)^2\pi^2\sigma_t^4}{r_t^5} -\frac{\sigma_t^2}{r_t^3} \bigg\} 
\exp\bigg\{ -\frac{(2n_d-1)^2\pi^2\sigma_t^2}{2r_t^2} \bigg\}.
\end{equation}

Summarizing the above discussion, I obtain the next proposition.

\vspace{0.5\baselineskip}

\noindent
{\bf Proposition 1.} \\
The probability density of $r_t$ given $\sigma^2_t$ is expressed as follows: \\
(a) (\cite{Feller(1951)})
\[
f_{\mathrm{range}}(r_t|\sigma^2_t) =8\sum_{n_d=1}^\infty (-1)^{n_d-1} \frac{n_d^2}{\sqrt{2\pi\sigma_t^2}}\exp\bigg( -\frac{n_d^2r_t^2}{2\sigma_t^2} \bigg).
\]
(b)
\[
f_{\mathrm{range}}(r_t|\sigma^2_t) =8\sum_{n_d=1}^\infty \bigg\{ \frac{(2n_d-1)^2\pi^2\sigma_t^4}{r_t^5} -\frac{\sigma_t^2}{r_t^3} \bigg\} 
\exp\bigg\{ -\frac{(2n_d-1)^2\pi^2\sigma_t^2}{2r_t^2} \bigg\}.
\]

\subsection{Generation of range}
In this subsection,  I consider the generation of a random variable with density stated in the last subsection.
I develop a random sampling algorithm for $x=r_t^2/\sigma_t^2$ using an alternating series method based on \cite{Devroye(1986)}.

Assume that the density function for a random variable $z$ is represented as $f(z)=ch(z)\sum_{i=0}^\infty a_i(z)$, 
where $c$ is a constant, 
a sequence of functions $\{a_i(z)\}_{i=0}^\infty$ ($a_0(z)\equiv 1$) is decreasing, 
and $h$ is a density function.
I also assume that it is easy to generate a random sample from $h$. 
Then, the following inequality for $f$ and the partial sums $A_j(z) =ch(z)\sum_{i=0}^j a_i(z), \; j=0, 1, \ldots,$ holds: 
\[
A_0(z) \ge A_2(z) \ge \cdots \ge f(z) \ge \cdots \ge A_3(z) \ge A_1(z).
\]
A random sample can be generated from density function $f$ using an acceptance-rejection algorithm as shown below. 
\begin{enumerate}
\item 
Generate $Z \sim h$.
\item 
Generate $U$ from uniform density on unit interval denoted by $\mathrm{U}(0,1)$. 
\item 
Calculate $A_i(Z), \; i=1, 2, \ldots$ iteratively until $A_j(Z)/\{cf(Z)\}\ge U$ for an odd $j$ or $A_j(Z)/\{cf(Z)\}< U$ for an even $j$.
\item 
If $j$ is odd, accept $Z$. If $j$ is even, return to Step 1.
\end{enumerate}

\vspace{0.5\baselineskip}

Notice that the density function for $x=r_t^2/\sigma_t^2$ is represented in two forms: 
\begin{equation} \label{eq:2.2-a}
f(x) =4(2\pi)^{-\frac{1}{2}}x^{-\frac{1}{2}}\exp\Big(-\frac{x}{2}\Big)\sum_{n_d=0}^\infty (-1)^{n_d} a_{n_d}, 
\end{equation}
where 
$a_{n_d} = (n_d+1)^2 \exp\{-\frac{1}{2}((n_d+1)^2-1)x\}$, 
and
\begin{equation} \label{eq:2.2-b}
f(x) =4\pi^2 x^{-3} \exp\bigg( -\frac{\pi^2}{2x} \bigg)\sum_{n_d=0}^\infty (-1)^{n_d} a_{n_d}, 
\end{equation}
where 
\begin{equation*}
a_{n_d} = \begin{cases}
\frac{x}{\pi^2} \exp\big\{-\frac{\pi^2(n_d^2-1)}{2x}\big\} & \text{ if } n_d \text{ is odd}, \\
(n_d+1)^2 \exp\big\{-\frac{\pi^2((n_d+1)^2-1)}{2x}\big\} & \text{ if } n_d \text{ is even}.
\end{cases}
\end{equation*}

I have the following lemma. (The proof is very similar to that of Lemma 5.1 in \cite{Devroye(1986)}.) 

\vspace{0.5\baselineskip}

\noindent
{\bf Lemma 1.} 

(a) $a_{n_d}$'s in equation (\ref{eq:2.2-a}) are monotone decreasing for $x>\frac{4}{3}$.

(b) $a_{n_d}$'s in equation (\ref{eq:2.2-b}) are monotone decreasing for $x<\pi^2$.

\vspace{0.25\baselineskip}

\noindent
{\bf Proof:} 

For the first series expansion (\ref{eq:2.2-a}), I have 
\begin{align}
\log\bigg(\frac{a_{n_d+1}}{a_{n_d}} \bigg) &= 2\log\frac{n_d+2}{n_d+1} -\frac{(2n_d+3)x}{2} \notag \\
&\le \frac{2}{n_d+1} -\frac{(2n_d+3)x}{2} \notag \\
&\le 2 -\frac{3x}{2} \notag \\
&< 0,
\end{align}
if $x>\frac{4}{3}$. 
In the second series expansion (\ref{eq:2.2-b}), for $n_d$ even and for $x<\pi^2$, I have 
\begin{equation}
\frac{a_{n_d}}{a_{n_d+1}} = \frac{(n_d+1)^2\pi^2}{x} \ge \frac{\pi^2}{x} >1, 
\end{equation}
and
\begin{align}
\log\bigg(\frac{a_{n_d+1}}{a_{n_d+2}} \bigg) &= -\log\frac{(n_d+2)^2\pi^2}{x} +\frac{(2n_d+3)\pi^2}{2x} \notag \\
&= -2\log(n_d+2) -\log\frac{\pi^2}{x}+\frac{(2n_d+3)\pi^2}{2x} \notag \\
&> -2\log(n_d+2) +\frac{2n_d+3}{2} \notag \\
&> 0.
\end{align}
\hspace{\fill}
$\Box$

\vspace{0.5\baselineskip}

This lemma implies that for at least one of the representations the condition stated below is satisfied: 
\[
A_0(x) \ge A_2(x) \ge \cdots \ge f(x) \ge \cdots \ge A_3(x) \ge A_1(x),
\]
where $x$ is real and positive. 
The lemma also shows that the inequality holds true for neither of the representations on the entire support of $f$. 
It suggests that I can use the alternating series method with 
\begin{equation}
a_{n_d} = \begin{cases}
(n_d+1)^2 \exp\{-\frac{1}{2}((n_d+1)^2-1)x\}  & \text{ if } x> c_{\mathrm{th}}, \\
\frac{x}{\pi^2} \exp\big\{-\frac{\pi^2(n_d^2-1)}{2x}\big\} & \text{ if } x\le c_{\mathrm{th}} \text{ and } n_d \text{ is odd}, \\
(n_d+1)^2 \exp\big\{-\frac{\pi^2((n_d+1)^2-1)}{2x}\big\} & \text{ if } x\le c_{\mathrm{th}} \text{ and } n_d \text{ is even},
\end{cases}
\end{equation}
and 
\begin{equation}
ch(x)= \begin{cases}
4(2\pi)^{-\frac{1}{2}}x^{-\frac{1}{2}}\exp\big(-\frac{x}{2}\big) & \text{ if } x> c_{\mathrm{th}}, \\
4\pi^2 x^{-3} \exp\big( -\frac{\pi^2}{2x} \big) & \text{ if } x\le c_{\mathrm{th}}, 
\end{cases}
\end{equation}
where the threshold $c_{\mathrm{th}}$ is in between the range $(\frac{4}{3}, \pi^2)$. 
Thus, I find that as a proposal $X$ may be sampled from a mixture distribution of chi squared and inverse gamma:  
\begin{equation}
X \sim \begin{cases}
\chi^2_1 \mathrm{I}(x>c_{th}) & \text{ w.p. } p/(p+q), \\
\mathrm{IG}(2,\pi^2/2) \mathrm{I}(x\le c_{th}) &  \text{ w.p. } q/(p+q), 
\end{cases}
\end{equation}
where 
$p =\int_{c_{\mathrm{th}}}^{\infty}  (2\pi)^{-\frac{1}{2}}x^{-\frac{1}{2}}\exp(-\frac{x}{2})dx$, 
$q =\int_{0}^{c_{\mathrm{th}}} \pi^2 x^{-3} \exp ( -\frac{\pi^2}{2x} ) dx $, 
and $\mathrm{I}(\cdot)$ is an indicator function.

Summarizing above, I obtain the algorithm that generates a random sample $x$ as follows: 

\begin{enumerate}
\item 
Generate $X \sim h(x)$ as a proposal.
\item 
Generate $U \sim \mathrm{U}(0,1)$.
\item 
Calculate $A_i(X), \; i=1, 2, \ldots$ iteratively until $A_j(X)/\{ch(X)\}\ge U$ for an odd $j$ or $A_j(X)/\{ch(X)\}< U$ for an even $j$.
\item 
If $j$ is odd, accept $X$. If $j$ is even, return to Step 1.
\end{enumerate}

\subsection{Bias correction for range}
Based on \cite{Feller(1951)}, \cite{Parkinson(1980)} showed that the $p$-th moment of daily return is represented as 
\begin{equation}
\mathrm{E}(r_t^p) =\frac{4}{\sqrt{\pi}} \Gamma\Big(\frac{p+1}{2}\Big) \Big( 1-\frac{4}{2^p} \Big) \zeta(p-1) (2\sigma^2_t)^{\frac{p}{2}}, 
\end{equation}
where $\Gamma(\cdot)$ denotes gamma function and $\zeta(\cdot)$ is Riemann zeta function. 
Note that $\mathrm{E}(r_t) =\sqrt{8\sigma^2_t/\pi}$ and $\mathrm{E}(r_t^2) =(4\log 2)\sigma^2_t$ and that $r_t^2/(4\log 2)$ is often used as a volatility estimator. 
Contrary to the assumption that the assets are driven by a Brownian motion and have a continuous price, the asset prices observed in the market take discrete values. 
In addition, the market is not open 24/7 for trading. 
Thus, several authors suggested that the observed high-low range $r_t$ is regarded as biased. 

In this paper, the high-low range $r_t$ observed at the market is assumed to be proportional to the true range $\widetilde{r}_t$ as
\begin{equation}
r_t =\lambda_t^{\frac{1}{2}} \widetilde{r}_t, \;\; \lambda_t \sim\mathrm{G}\Big(\frac{\nu_1}{2}, \frac{\nu_2}{2}\Big), \;\; t=1, \ldots, n,  
\end{equation}
where a gamma distribution is denoted as $\mathrm{G}(\alpha, \beta)$. 
If the scaling parameter $\lambda_t$ is smaller than $1$, it implies that the observed range $r_t$ has a downward bias.

\subsection{The model}
The asymmetric SV model is represented using state space form as
\begin{align}
y_t &=\sigma_t\epsilon_t, \;\; t=1, \ldots, n, \label{eq:2.4-a} \\
\log(\sigma_{t+1}^2) &= \phi\log(\sigma_t^2) +\eta_t,  \;\; t=1, \ldots, n-1, \\
\begin{pmatrix} \epsilon_t \\ \eta_t \end{pmatrix} 
&\sim \mathrm{N}(\bm{0}_2, \Omega), \;\; 
\Omega =\begin{pmatrix} 1 & \omega_{\epsilon\eta} \\ \omega_{\eta\epsilon} & \omega_{\eta\eta} \end{pmatrix}, \\
\log(\sigma_{1}^2) &= \eta_1, \;\; \eta_1 \sim\mathrm{N}(0, \omega_{\eta\eta, 0}), \;\; \omega_{\eta\eta, 0} =\omega^2/(1-\phi^2), 
\end{align}
where $y_t$ is a daily asset return, $\sigma^2_t$ is the unobserved volatility, 
$\mathrm{N}(\bm{m}, S)$ denotes a normal distribution with mean vector $\bm{m}$ and covariance matrix $S$, $\bm{0}_p$ is a $p$-dimensional vector of zeros and $|\phi|<1$. 
If $\omega_{\eta\epsilon}$ takes a negative value, it means the existence of asymmetry or leverage effect as stated in Section 1. 

I add the following observation equation to the state space model described above: 
\begin{equation} \label{eq:2.4-b} 
r_t =\lambda_t^{\frac{1}{2}} \widetilde{r}_t, \;\; \widetilde{r}_t \sim f_{\mathrm{range}}(\widetilde{r}_t|\sigma^2_t), \;\; \lambda_t ~\sim\mathrm{G}\Big(\frac{\nu_1}{2}, \frac{\nu_2}{2}\Big), \;\; t=1, \ldots, n, 
\end{equation}
where $f_{\mathrm{range}}$ is the density function defined in Proposition 1. 
The above Equations (\ref{eq:2.4-a})--(\ref{eq:2.4-b}) define the stochastic volatility model with range-based correction and leverage (SVRG). 

Let $\widetilde{\sigma}^2_t =\lambda_t\sigma^2_t$ and notice that $r_t \sim f_{\mathrm{range}}(r_t |\widetilde{\sigma}^2_t)$. 
Thus, I may rewrite the proposed SVRG model as follows: 
\begin{align}
y_t &=\lambda_t^{-\frac{1}{2}}\widetilde{\sigma}_t\epsilon_t, \;\; t=1, \ldots, n, \\
r_t &\sim f_{\mathrm{range}}(r_t |\widetilde{\sigma}^2_t), \;\; t=1, \ldots, n, \\
\log(\widetilde{\sigma}_{t+1}^2) &= \log\lambda_{t+1} +\phi\{\log(\widetilde{\sigma}_t^2) -\log\lambda_t\} +\eta_t,  \;\; t=1, \ldots, n-1, \\
\log(\widetilde{\sigma}_{1}^2) &= \log\lambda_1 +\eta_1, \;\; \eta_1 \sim\mathrm{N}(0, \omega_{\eta\eta, 0}). 
\end{align}

\section{Bayesian estimation}
\subsection{Assumption for prior}
I assume the prior of $\phi$ that 
\begin{equation}
\frac{\phi +1}{2} \sim\mathrm{Be}(a_{\phi 0}, b_{\phi 0}), 
\end{equation}
where a beta distribution with parameters $(a,b)$ is denoted as $\mathrm{Be}(a,b)$. 

For $\Omega$, let  
\begin{equation}
\Omega =\begin{pmatrix}
\omega_{\epsilon\epsilon} & \omega_{\epsilon\eta} \\
\omega_{\eta\epsilon} & \omega_{\eta\eta}
\end{pmatrix}, \;\;
\Omega^{-1} =\begin{pmatrix}
\omega^{(\epsilon\epsilon)} & \omega^{(\epsilon\eta)} \\
\omega^{(\eta\epsilon)} & \omega^{(\eta\eta)}
\end{pmatrix},
\end{equation}
and notice that 
$\omega^{(\epsilon\epsilon)} =1 +(\omega^{(\epsilon\eta)})^2 (\omega^{(\eta\eta)})^{-1}$. 
Thus, I assume the prior distribution for $\Omega^{-1}$ such that 
\begin{equation}
\omega^{(\eta\eta)}\sim\mathrm{G}\bigg(\frac{n_0}{2}, \frac{1}{2s_0}\bigg), \;\; 
\omega^{(\epsilon\eta)}|\omega^{(\eta\eta)}\sim\mathrm{N}(\omega^{(\eta\eta)}\delta_0, \gamma_0\omega^{(\eta\eta)}). 
\end{equation}

Let $\bm{\nu}=(\nu_1, \nu_2)'$ and an assumption for the prior of $\bm{\nu}$ is made as follows: 
\begin{equation}
\nu_1 \sim\mathrm{G}\bigg( \frac{\alpha_{\nu_1 0}}{2}, \frac{\beta_{\nu_1 0}}{2} \bigg), \;\;
\nu_2 \sim\mathrm{G}\bigg( \frac{\alpha_{\nu_2 0}}{2}, \frac{\beta_{\nu_2 0}}{2} \bigg). 
\end{equation}

Define $\bm{\vartheta}=(\phi, \Omega, \bm{\nu})$ and let us denote the prior probability density of $\bm{\vartheta}$ as $\pi(\bm{\vartheta})$. 
I obtain the joint posterior probability density that is expressed as 
\begin{align}
&f(\bm{\vartheta}, \{\sigma^2_t\}_{t=1}^n, \{\lambda_t\}_{t=1}^n | \{y_t\}_{t=1}^n, \{r_t\}_{t=1}^n) \notag \\
&\propto \pi(\bm{\vartheta}) f(\sigma^2_1 |\bm{\vartheta}) \prod_{t=1}^{n} f(y_t |\sigma^2_{t}, \bm{\vartheta}) 
 \prod_{t=1}^{n-1} f(\sigma^2_{t+1} |y_t, \sigma^2_{t}, \bm{\vartheta}) \prod_{t=1}^{n} f(r_t |\sigma^2_t, \lambda_t) \prod_{t=1}^{n} f(\lambda_t |\bm{\vartheta}) \\
&\propto \pi(\bm{\vartheta}) f(\widetilde{\sigma}^2_1 |\lambda_t, \bm{\vartheta}) \prod_{t=1}^{n} f(y_t |\widetilde{\sigma}^2_{t}, \lambda_t, \bm{\vartheta}) 
 \prod_{t=1}^{n-1} f(\widetilde{\sigma}^2_{t+1} |y_t, \lambda_{t+1}, \widetilde{\sigma}^2_{t}, \lambda_t, \bm{\vartheta}) \notag \\
&\hspace{15pt} \times\prod_{t=1}^{n} f_{\mathrm{range}}(r_t |\widetilde{\sigma}^2_t) \prod_{t=1}^{n} f(\lambda_t |\bm{\vartheta}), 
\end{align}
where
\begin{align}
&f(y_t |\sigma^2_{t}, \bm{\vartheta}) \propto (\sigma^2_t)^{-\frac{1}{2}} \exp\bigg(-\frac{y_t^2}{2\sigma^2_t} \bigg), \\
&f(\sigma^2_{t+1} |y_t, \sigma^2_{t}, \bm{\vartheta}) \propto (\sigma^2_{t+1})^{-1} (\omega^{(\eta\eta)})^{\frac{1}{2}} \notag \\
&\hspace{20pt} \times \exp\bigg[ -\frac{1}{2} \omega^{(\eta\eta)}
\{\log\sigma^2_{t+1} -\phi\log\sigma^2_t -\omega_{\eta\epsilon}(\sigma^2_t)^{-\frac{1}{2}} y_t \}^2
\bigg], \\
&f(\sigma^2_1 |\bm{\vartheta}) \propto (\sigma^2_1)^{-1} ( \omega_{\eta\eta, 0})^{-\frac{1}{2}} 
 \exp\bigg\{ -\frac{(\log\sigma^2_1 )^2}{2\omega_{\eta\eta, 0}} \bigg\}, \\
&f(y_t |\widetilde{\sigma}^2_{t}, \lambda_t, \bm{\vartheta}) \propto \lambda_t^{\frac{1}{2}} (\widetilde{\sigma}^2_t)^{-\frac{1}{2}} \exp\bigg(-\frac{y_t^2\lambda_t}{2\widetilde{\sigma}^2_t} \bigg), \\
&f(\widetilde{\sigma}^2_{t+1} |y_t, \lambda_{t+1}, \widetilde{\sigma}^2_{t}, \lambda_t, \bm{\vartheta}) 
\propto (\widetilde{\sigma}^2_{t+1})^{-1} (\omega^{(\eta\eta)})^{\frac{1}{2}} \notag \\
&\hspace{20pt} \times\exp\bigg[ -\frac{1}{2} \omega^{(\eta\eta)}
\{\log\widetilde{\sigma}^2_{t+1} -\log\lambda_{t+1} -\phi(\log\widetilde{\sigma}^2_t -\log\lambda_t) -\omega_{\eta\epsilon}(\widetilde{\sigma}^2_t)^{-\frac{1}{2}} \lambda_t^{\frac{1}{2}} y_t \}^2
\bigg], \\
&f(\widetilde{\sigma}^2_1 |\lambda_1, \bm{\vartheta}) \propto (\widetilde{\sigma}^2_1)^{-1} ( \omega_{\eta\eta, 0})^{-\frac{1}{2}} 
 \exp\bigg\{ -\frac{(\log\widetilde{\sigma}^2_1 -\log\lambda_1)^2}{2\omega_{\eta\eta, 0}} \bigg\}, \\
&f(\lambda_t |\bm{\vartheta}) =
\frac{(\frac{\nu_2}{2})^{\frac{\nu_1}{2}}}{\Gamma(\frac{\nu_1}{2})} \lambda_t^{\frac{\nu_1}{2}-1} \exp\bigg( -\frac{\nu_2}{2} \lambda_t \bigg).  
\end{align}

The MCMC algorithm is implemented in five blocks: 
\begin{enumerate}
\item 
Initialize $\{\sigma^2_t\}_{t=1}^n, \; \{\lambda_t\}_{t=1}^n, \; \bm{\vartheta}$.
\item 
Generate $\{\sigma^2_t\}_{t=1}^n |\{y_t\}_{t=1}^n, \{r_t\}_{t=1}^n, \{\lambda_t\}_{t=1}^n, \bm{\vartheta}$.
\item 
Generate $\{\lambda_t\}_{t=1}^n |\{y_t\}_{t=1}^n, \{\widetilde{\sigma}^2_t\}_{t=1}^n, \bm{\vartheta}$.
\item 
Generate $\bm{\vartheta} |\{y_t\}_{t=1}^n, \{r_t\}_{t=1}^n, \{\sigma^2_t\}_{t=1}^n, \{\lambda_t\}_{t=1}^n$.
\item 
Go to 2.
\end{enumerate}

\subsection{Generation of $\{\sigma^2_t\}_{t=1}^n$}
The conditional posterior density function of $\sigma^2_t$ is given by
\begin{align}
&f(\sigma^2_{t} |\{y_s\}_{s=1}^n, \{r_s\}_{s=1}^n, \{\sigma^2_{s}\}_{s\neq t}, \{\lambda_s\}_{s=1}^n, \bm{\vartheta}) \notag \\
&\propto \begin{cases}
f(\sigma^2_{t+1} |y_t, \sigma^2_{t}, \bm{\vartheta}) f(\sigma^2_{t} |\bm{\vartheta}) 
f(y_t |\sigma^2_{t}, \bm{\vartheta}) f(r_t |\sigma^2_t, \lambda_t)
& \text{ for } t=1, \\
f(\sigma^2_{t+1} |y_t, \sigma^2_{t}, \bm{\vartheta}) f(\sigma^2_{t} |y_{t-1}, \sigma^2_{t-1}, \bm{\vartheta}) 
f(y_t |\sigma^2_{t}, \bm{\vartheta}) f(r_t |\sigma^2_t, \lambda_t) 
& \text{ for } t=2, \ldots, n-1, \\
f(\sigma^2_{t} |y_{t-1}, \sigma^2_{t-1}, \bm{\vartheta}) 
f(y_t |\sigma^2_{t}, \bm{\vartheta}) f(r_t |\sigma^2_t, \lambda_t) 
& \text{ for } t=n. \end{cases}
\end{align}

Let $h_{\sigma^2_t} =\log\sigma^2_t$. Using Taylor expansion around, say, $c_{\sigma^2_t}$, I obtain an approximation for $\sigma_t^{-1}$ as
\begin{align}
\sigma_t^{-1} =\exp\Big(-\frac{h_{\sigma^2_t}}{2}\Big) &\approx \exp\Big(-\frac{c_{\sigma^2_t}}{2}\Big) -\frac{1}{2}\exp\Big(-\frac{c_{\sigma^2_t}}{2}\Big)(h_{\sigma^2_t}-c_{\sigma^2_t}) \notag \\ &=\exp\Big(-\frac{c_{\sigma^2_t}}{2}\Big)\Big(1+\frac{c_{\sigma^2_t}}{2}\Big) -\frac{1}{2}\exp\Big(-\frac{c_{\sigma^2_t}}{2}\Big)\log(\sigma^2_t).
\end{align}
It is recommended to set $c_{\sigma^2_t} =\log\{ r_t^2/(4\lambda_t\log 2) \}$. 
For $t=1, \ldots, n-1,$ I may consider an approximation of $f(\sigma^2_{t+1} |y_t, \sigma^2_{t}, \bm{\vartheta})$ as 
\begin{equation}
f(\sigma^2_{t+1} |y_t, \sigma^2_{t}, \bm{\vartheta}) 
\approx \exp \bigg\{ -\frac{1}{2s_{\sigma^2_t, 1}}(\log\sigma^2_t -m_{\sigma^2_t, 1})^2 \bigg\} \times\text{ constant}, 
\end{equation}
where 
\begin{align}
&s_{\sigma^2_t, 1} =\bigg[ \bigg\{ \phi-\omega_{\eta\epsilon}\cdot\frac{1}{2}\exp\Big(-\frac{1}{2}c_{\sigma^2_t}\Big) y_t \bigg\}^2 \omega^{(\eta\eta)} \bigg]^{-1}, \\
&m_{\sigma^2_t, 1} 
= 
\bigg\{ \phi-\omega_{\eta\epsilon} \cdot\frac{1}{2}\exp\Big(-\frac{1}{2}c_{\sigma^2_t}\Big) y_t \bigg\}^{-1} 
\bigg\{\log\sigma^2_{t+1} -\omega_{\eta\epsilon} \Big(1+\frac{1}{2}c_{\sigma^2_t}\Big) \exp\Big(-\frac{1}{2}c_{\sigma^2_t}\Big) y_t \bigg\}. 
\end{align}

Consider the following function,  
\begin{equation}
g_{\sigma^2_t}(\sigma^2_{t}) 
= \begin{cases}
f(\sigma^2_{t+1} |y_t, \sigma^2_{t}, \bm{\vartheta})f(\sigma^2_{t} |\bm{\vartheta})
& \text{ for } t=1, \\
f(\sigma^2_{t+1} |y_t, \sigma^2_{t}, \bm{\vartheta})f(\sigma^2_{t} |y_{t-1}, \sigma^2_{t-1}, \bm{\vartheta})
& \text{ for } t=2, \ldots, n-1, \\
f(\sigma^2_{t} |y_{t-1}, \sigma^2_{t-1}, \bm{\vartheta}) 
& \text{ for } t=n, \end{cases}
\end{equation}
and note that $g_{\sigma^2_t}$ is approximated by a function proportional to the probability density of log-normal distribution as shown below: 
\begin{align}
g_{\sigma^2_t}(\sigma^2_{t} ) 
&\approx (\sigma^2_t)^{-1} \exp \bigg\{ -\frac{1}{2s_{\sigma^2_t}}(\log\sigma^2_t -m_{\sigma^2_t})^2 \bigg\} \times\text{constant}, \\
s_{\sigma^2_t} &=\begin{cases}
(s_{\sigma^2_t, 1}^{-1} +\omega_{\eta\eta,0}^{-1})^{-1} & \text{ if } t=1, \\
(s_{\sigma^2_t, 1}^{-1} +\omega^{(\eta\eta)})^{-1} & \text{ if } t=2, \ldots, n-1, \\
\omega_{\eta\eta} -\omega_{\eta\epsilon}^2 & \text{ if } t=n, 
\end{cases} \\
m_{\sigma^2_t} &=\begin{cases}
s_{\sigma^2_t}(s_{\sigma^2_t,1}^{-1}m_{\sigma^2_t,1} +\omega_{\eta\eta,0}^{-1}m_{\sigma^2_t,2}) & \text{ if } t=1, \\
s_{\sigma^2_t}(s_{\sigma^2_t,1}^{-1}m_{\sigma^2_t,1} +\omega^{(\eta\eta)}m_{\sigma^2_t,2}) & \text{ if } t=2, \ldots, n-1, \\
m_{\sigma^2_t,2} & \text{ if } t=n, 
\end{cases} \\
m_{\sigma^2_t, 2} &=\begin{cases}
0 & \text{ if } t=1, \\
 \phi (\log\sigma^2_{t-1}) +\omega_{\eta\epsilon} y_{t-1}/\sigma_{t-1} & \text{ if } t=2, \ldots, n. 
\end{cases}
\end{align}

Let $\alpha_{\sigma^2_t} =(\exp(s_{\sigma^2_t})-1)^{-1}+2$, and $\beta_{\sigma^2_t} =\exp(m_{\sigma^2_t}+\frac{1}{2}s_{\sigma^2_t}) \{(\exp(s_{\sigma^2_t})-1)^{-1}+1\}$. 
Notice that a random variable with the inverse gamma density function with parameters $(\alpha_{\sigma^2_t}, \beta_{\sigma^2_t})$ 
has mean $m_{\sigma^2_t}$ and variance $s_{\sigma^2_t}$.
It implies that the log-normal probability density function with (log) mean $m_{\sigma^2_t}$ and (log) variance $s_{\sigma^2_t}$ is approximated by inverse gamma distribution with parameters $(\alpha_{\sigma^2_t}, \beta_{\sigma^2_t})$.

Then, an approximation of $f(\sigma^2_{t} |\{y_s\}_{s=1}^n, \{r_s\}_{s=1}^n, \{\sigma^2_{s}\}_{s\neq t}, \{\lambda_s\}_{s=1}^n, \bm{\vartheta})$ is given by the probability density function 
$f_a(\sigma^2_{t} |\{y_s\}_{s=1}^n, \{\widetilde{r}_s\}_{s=1}^n, \{\sigma^2_{s}\}_{s\neq t}, \bm{\vartheta})$, where
\begin{align}
&f_a(\sigma^2_{t} |\{y_s\}_{s=1}^n, \{\widetilde{r}_s\}_{s=1}^n, \{\sigma^2_{s}\}_{s\neq t}, \bm{\vartheta}) \notag \\
&\propto 
(\sigma^2_t)^{-\alpha_{\sigma^2_t}-1}\exp\Big( -\frac{\beta_{\sigma^2_t}}{\sigma^2_t} \Big)
 \sum_{n_d=1}^\infty (-1)^{n_d-1} \frac{n_d^2}{\sqrt{2\pi}}(\sigma_t^2)^{
-1}
\exp\bigg( -\frac{n_d^2\widetilde{r}_t^2 +y_t^2 }{2\sigma_t^2} \bigg) \\
&=(2\pi)^{-\frac{1}{2}} (\sigma^2_t)^{-\nu_{t,1}-1} \exp\Big(-\frac{\delta_{t,1}}{2\sigma^2_t} \Big) \notag \\
&\hspace{20pt} \times \sum_{n_d=0}^\infty (-1)^{n_d-1} (n_d+1)^2
\exp\bigg\{ -\frac{((n_d+1)^2-1)\widetilde{r}_t^2}{2\sigma_t^2} \bigg\},  \label{eq:3.2-c}
\end{align}
and where
\begin{equation}
\nu_{t,1}= \alpha_{\sigma^2_t}+1, \;\; \delta_{t, 1} =\widetilde{r}_t^2 +y_t^2 +2\beta_{\sigma^2_t}. 
\end{equation}
The right hand of equation (\ref{eq:3.2-c}) is rewritten as
\begin{align}
&(\sigma^2_t)^{-\alpha_{\sigma^2_t}-1}\exp\bigg( -\frac{\beta_{\sigma^2_t}}{\sigma^2_t} \bigg) \notag \\
&\hspace{20pt} \times (\sigma_t^2)^{-\frac{1}{2}}
 \sum_{n_d=1}^\infty \bigg\{ \frac{(2n_d-1)^2\pi^2\sigma_t^4}{\widetilde{r}_t^5} -\frac{\sigma_t^2}{\widetilde{r}_t^3} \bigg\} 
\exp\bigg\{ -\frac{(2n_d-1)^2\pi^2\sigma_t^2}{2\widetilde{r}_t^2} -\frac{1}{2} \sigma_t^{-2} y_t^2\bigg\} \notag \\
&= \frac{\pi^2}{\widetilde{r}_t^5} 
(\sigma^2_t)^{\nu_{t,2}-1} \exp\bigg\{ -\frac{1}{2}\Big(\frac{\delta_{t,2}^2}{\sigma^2_t} +\gamma_{t, 2}^2\sigma^2_t \Big)\bigg\} \notag \\
&\hspace{20pt} \times \sum_{n_d=1}^\infty \bigg\{ 
(2n_d-1)^2 -\frac{\widetilde{r}_t^2}{\pi^2\sigma_t^2} \bigg\} 
\exp\bigg\{ -\frac{((2n_d-1)^2-1)\pi^2\sigma_t^2}{2\widetilde{r}_t^2} \bigg\}, 
\end{align}
where
\begin{equation}
\nu_{t,2}= -\alpha_{\sigma^2_t}+\frac{3}{2}, \;\; \delta_{t, 2} =(y_t^2 +2\beta_{\sigma^2_t})^{\frac{1}{2}}, \;\; \gamma_{t,2} =\frac{\pi}{\widetilde{r}_t}.  
\end{equation}
I notice that $f_a(\sigma^2_t|\cdot)$ is represented using a constant $c$, an easy density function $h$ and a series $b_{n_d}$ as $f_a(\sigma^2_t|\cdot)=ch(\sigma^2_t|\cdot)\sum_{n_d=0}^\infty (-1)^{n_d} b_{n_d}$, where 
\begin{equation} \label{eq:3.2-a}
ch(\sigma^2_t|\cdot) =(2\pi)^{-\frac{1}{2}} (\sigma^2_t)^{-\nu_{t,1}-1} \exp\Big(-\frac{\delta_{t,1}}{2\sigma^2_t} \Big), \;\; 
b_{n_d} =(n_d+1)^2 \exp\bigg\{ -\frac{((n_d+1)^2-1)\widetilde{r}_t^2}{2\sigma_t^2} \bigg\}, 
\end{equation}
or 
\begin{align}
&ch(\sigma^2_t|\cdot) = \frac{\pi^2}{\widetilde{r}_t^5} 
(\sigma^2_t)^{\nu_{t,2}-1} \exp\bigg\{ -\frac{1}{2}\Big(\frac{\delta_{t,2}^2}{\sigma^2_t} +\gamma_{t, 2}^2\sigma^2_t \Big)\bigg\},  \label{eq:3.2-b} \\
&b_{n_d} =
\begin{cases}
\frac{\widetilde{r}_t^2}{\pi^2\sigma_t^{2}} \exp\bigg\{ -\frac{(n_d^2-1)\pi^2\sigma_t^2}{2\widetilde{r}_t^2} \bigg\}, & (n_d\text{: odd}) \\
(n_d+1)^2 \exp\bigg\{ -\frac{((n_d+1)^2-1)\pi^2\sigma_t^2}{2\widetilde{r}_t^2} \bigg\}, & (n_d\text{: even}) \\
\end{cases}.
\end{align}

I have the following lemma.

\vspace{\baselineskip}

\noindent
{\bf Lemma 2.} 

(a) $b_{n_d}$'s in equation (\ref{eq:3.2-a}) are monotone decreasing for $\sigma^2_t<\frac{3}{4}\widetilde{r}_t^2$.

(b) $b_{n_d}$'s in equation (\ref{eq:3.2-b}) are monotone decreasing for $\sigma^2_t>\frac{\widetilde{r}_t^2}{\pi^2}$.

\vspace{\baselineskip}

I may prove this lemma exactly the same as Lemma 1.

This lemma implies that for at least one of the representations the inequality stated below holds true: 
\[
B_0(\sigma^2_t) \ge B_2(\sigma^2_t) \ge \cdots \ge f_a(\sigma^2_t) \ge \cdots \ge B_3(\sigma^2_t) \ge B_1(\sigma^2_t),
\]
where $B_j(z) =ch(z)\sum_{i=0}^j b_i(z)$. 
The lemma also shows that for neither of the representations the inequality holds true on the entire support of $f_a$. 
This suggests that the alternating series method can be used with 
\begin{equation}
b_{n_d} = \begin{cases}
(n_d+1)^2 \exp\{-\frac{((n_d+1)^2-1)\widetilde{r}_t^2}{2\sigma_t^2}\}  & \text{ if } \sigma^2_t <c_{\mathrm{th}}\widetilde{r}_t^2, \\
\frac{\widetilde{r}_t^2}{\pi^2\sigma_t^{2}} 
\exp\bigg\{ -\frac{(n_d^2-1)\pi^2\sigma_t^2}{2\widetilde{r}_t^2} \bigg\} & \text{ if } \sigma^2_t \ge c_{\mathrm{th}}\widetilde{r}_t^2 \text{ and } (n_d\text{: odd}) \\
(n_d+1)^2 \exp\bigg\{ -\frac{((n_d+1)^2-1)\pi^2\sigma_t^2}{2\widetilde{r}_t^2} \bigg\} & \text{ if } \sigma^2_t \ge c_{\mathrm{th}}\widetilde{r}_t^2 \text{ and } (n_d\text{: even}), 
\end{cases}
\end{equation}
and 
\begin{equation}
ch(x)= \begin{cases}
(2\pi)^{-\frac{1}{2}} (\sigma^2_t)^{-\nu_{t,1}-1} \exp\Big(-\frac{\delta_{t,1}}{2\sigma^2_t} \Big)  & \text{ if } \sigma^2_t <c_{\mathrm{th}}\widetilde{r}_t^2, \\
\frac{\pi^2}{\widetilde{r}_t^5} (\sigma^2_t)^{\nu_{t,2}-1} \exp\bigg\{ -\frac{1}{2}\Big(\frac{\delta_{t,2}^2}{\sigma^2_t} +\gamma_{t, 2}^2\sigma^2_t \Big)\bigg\} & \text{ if } \sigma^2_t \ge c_{\mathrm{th}}\widetilde{r}_t^2, 
\end{cases}
\end{equation}
where the threshold $c_{\mathrm{th}}$ is in between the range $(\pi^{-2}\widetilde{r}_t^2, \frac{3}{4}\widetilde{r}_t^2)$. 
Thus, I find that as a proposal $\sigma^2_t$ may be sampled from a mixture distribution of inverse gamma and generalized inverse Gaussian:  
\begin{equation}
\sigma^2_t \sim \begin{cases}
\mathrm{IG}(\nu_{t,1}, \delta_{t,1}/2) \mathrm{I}(\sigma_t^2 < c_{\mathrm{th}}\widetilde{r}_t^2) & \text{ w.p. } p_t/(p_t+q_t), \\
\mathrm{GIG}(\nu_{t,2}, \delta_{t,2}, \gamma_{t,2}) \mathrm{I}(\sigma_t^2 \ge c_{\mathrm{th}}\widetilde{r}_t^2) & \text{ w.p. } q_t/(p_t+q_t). 
\end{cases}
\end{equation}
Note that 
\begin{equation}
p_t =\int_{0}^{c_{\mathrm{th}}\widetilde{r}_t^2} (2\pi)^{-\frac{1}{2}} (\sigma^2_t)^{-\nu_{t,1}-1} \exp\bigg(-\frac{\delta_{t,1}}{2\sigma^2_t} \bigg) d\sigma^2_t  
=\frac{1}{\sqrt{2\pi} (\delta_{t,1}/2)^{\nu_{t,1}}} \Gamma \bigg( \nu_{t,1}, \frac{\delta_{t,1}}{2c_{\mathrm{th}} \widetilde{r}_t^2}\bigg),
\end{equation}
where $\Gamma(\alpha, x)$ is an incomplete gamma function defined as 
\begin{equation}
\Gamma(\alpha, x) =\int_{x}^{\infty} t^{\alpha-1} e^{-t} dt,
\end{equation}
and 
\begin{align}
q_t &=\frac{\pi^2}{\widetilde{r}_t^5} \int_{c_{\mathrm{th}}\widetilde{r}_t^2}^{\infty}  (\sigma^2_t)^{\nu_{t,2}-1} \exp\bigg\{ -\frac{1}{2}\bigg(\frac{\delta_{t,2}^2}{\sigma^2_t} +\gamma_{t, 2}^2\sigma^2_t \bigg)\bigg\}d\sigma^2_t \notag \\
&=\frac{\pi^2}{\widetilde{r}_t^5} (c_{\mathrm{th}} \widetilde{r}_t^2)^{\nu_{t,2}} K_{-\nu_{t,2}} \bigg(\frac{c_{\mathrm{th}} \widetilde{r}_t^2 \gamma_{t,2}^2}{2}, \frac{\delta_{t,2}^2}{2c_{\mathrm{th}} \widetilde{r}_t^2} \bigg),
\end{align}
where $K_\nu(x,y)$ denotes an incomplete Bessel function defined as 
\begin{equation}
K_\nu(x,y) =\int_{1}^\infty \frac{e^{-xt -y/t}}{t^{\nu+1}} dt.
\end{equation}
A generalized inverse Gaussian distribution with parameters $(\nu, \delta, \gamma)$ is denoted as $\mathrm{GIG}(\nu, \delta, \gamma)$. 
These incomplete integrals appeared in $p_t$ and $q_t$ are calculated numerically (see \cite{Harris(2008)}).
An efficient sampling algorithm from a truncated gamma distribution is available from \cite{Philippe(1997)}. 
I generate a sample from a (truncated) generalized inverse Gaussian density using the algorithm proposed by \cite{Dagpunar(1989)}.

Now an algorithm for generating $\sigma^2_t$ is obtained as follows: 
\begin{enumerate}
\item 
Generate $(\sigma^2_t)^\dagger \sim h$ as a proposal.
\item 
Generate $U \sim \mathrm{U}(0,1)$.
\item 
Calculate $B_i((\sigma^2_t)^\dagger), \; i=1, 2, \ldots$ iteratively until $B_j((\sigma^2_t)^\dagger)/\{ch((\sigma^2_t)^\dagger)\}\ge U$ for an odd $j$ or $B_j((\sigma^2_t)^\dagger)/\{ch((\sigma^2_t)^\dagger)\}< U$ for an even $j$.
\item 
Accept $(\sigma^2_t)^\dagger$ as a proposal if $n_d$ is odd. Else return to step 1.
\item 
Accept $(\sigma^2_t)^\dagger$ with probability 
\begin{equation}
\min\left\{ 1, \frac{g_{\sigma^2_t}((\sigma^2_t)^\dagger) \cdot (\sigma^2_t)^{-\alpha_{\sigma^2_t}-1}\exp(-\beta_{\sigma^2_t}/\sigma^2_t)}{g_{\sigma^2_t}(\sigma^2_t) \cdot ((\sigma^2_t)^\dagger)^{-\alpha_{\sigma^2_t}-1}\exp(-\beta_{\sigma^2_t}/(\sigma^2_t)^\dagger)} \right\},
\end{equation}
where $\sigma^2_t$ is a current MCMC sample. 
\end{enumerate}

\subsection{Generation of $\{\lambda_t\}_{t=1}^n$}
The conditional posterior density function of $\lambda_t$ is derived as
\begin{align}
&f(\lambda_{t} |\{y_s\}_{s=1}^n, \{r_s\}_{s=1}^n, \{\widetilde{\sigma}^2_{s}\}_{s=1}^n, \{\lambda_s\}_{s\neq t}, \bm{\vartheta}) \notag \\
&\propto \begin{cases}
f(\lambda_t|\bm{\vartheta}) f(y_t |\widetilde{\sigma}^2_{t}, \lambda_t, \bm{\vartheta}) \\
\hspace{20pt} \times 
f(\widetilde{\sigma}^2_{t+1} |y_t, \lambda_{t+1}, \widetilde{\sigma}^2_{t}, \lambda_t, \bm{\vartheta}) f(\widetilde{\sigma}^2_{t} |\lambda_t, \bm{\vartheta}) 
& \text{ for } t=1, \\
f(\lambda_t|\bm{\vartheta}) f(y_t |\widetilde{\sigma}^2_{t}, \lambda_t, \bm{\vartheta}) \\
\hspace{20pt} \times f(\widetilde{\sigma}^2_{t+1} |y_t, \lambda_{t+1}, \widetilde{\sigma}^2_{t}, \lambda_t, \bm{\vartheta}) f(\widetilde{\sigma}^2_{t} |y_{t-1}, \lambda_t, \widetilde{\sigma}^2_{t-1}, \lambda_{t-1}, \bm{\vartheta}) 
& \text{ for } t=2, \ldots, n-1, \\
f(\lambda_t|\bm{\vartheta}) f(y_t |\widetilde{\sigma}^2_{t}, \lambda_t, \bm{\vartheta}) \\
\hspace{20pt} \times f(\widetilde{\sigma}^2_{t} |y_{t-1}, \lambda_{t}, \widetilde{\sigma}^2_{t-1}, \lambda_{t-1}, \bm{\vartheta}) 
& \text{ for } t=n. \end{cases}
\end{align}

Let us denote $\log\lambda_t$ by $h_{\lambda_t}$. 
I derive the Taylor expansion of $\lambda_t^{\frac{1}{2}}$ with respect to $h_{\lambda_t}$ around, say, $c_{\lambda_t}$, and obtain the approximation as  
\begin{align}
\lambda_t^{\frac{1}{2}} =\exp\Big(\frac{h_{\lambda_t}}{2}\Big) &\approx \exp\Big(\frac{c_{\lambda_t}}{2}\Big) +\frac{1}{2}\exp\Big(\frac{c_{\lambda_t}}{2}\Big)(h_{\lambda_t}-c_{\lambda_t}) \notag \\
&=\exp\Big(\frac{c_{\lambda_t}}{2}\Big)\Big(1-\frac{c_{\lambda_t}}{2}\Big) +\frac{1}{2}\exp\Big(\frac{c_{\lambda_t}}{2}\Big)\log(\lambda_t).
\end{align}
It is recommended to use the prior mode $\log\{ (\frac{\nu_1}{2}-1)/(\frac{\nu_2}{2}) \}$ as $c_{\lambda_t}$. 
Thus, for $t=1, \ldots, n-1,$ an approximation of $f(\widetilde{\sigma}^2_{t+1} |y_t, \lambda_{t+1}, \widetilde{\sigma}^2_{t}, \lambda_t, \bm{\vartheta})$ may be considered as 
\begin{align}
&f(\widetilde{\sigma}^2_{t+1} |y_t, \lambda_{t+1}, \widetilde{\sigma}^2_{t}, \lambda_t, \bm{\vartheta}) 
\approx \exp \bigg\{ -\frac{1}{2s_{\lambda_t, 1}}(\log\lambda_t -m_{\lambda_t, 1})^2 \bigg\} \times\text{constant}, \\
&s_{\lambda_t, 1} =\bigg[ \bigg\{ \phi-\omega_{\eta\epsilon}\cdot\frac{1}{2}\exp\Big(\frac{1}{2}c_{\lambda_t}\Big) \frac{y_t}{\widetilde{\sigma}_{t}} \bigg\}^2 \omega^{(\eta\eta)} \bigg]^{-1}, \\
&m_{\lambda_t, 1} 
= 
\bigg\{ \phi-\omega_{\eta\epsilon}\cdot\frac{1}{2}\exp\Big(\frac{1}{2}c_{\lambda_t}\Big) \frac{y_t}{\widetilde{\sigma}_{t}} \bigg\}^{-1} \notag \\
&\hspace{20pt} \times\bigg[ -\bigg\{\log\widetilde{\sigma}^2_{t+1} -\log\lambda_{t+1} -\phi\log\widetilde{\sigma}^2_{t}  -\omega_{\eta\epsilon} \Big(1-\frac{1}{2}c_{\lambda_t}\Big) \exp\Big(\frac{1}{2}c_{\lambda_t}\Big) \frac{y_t}{\widetilde{\sigma}_{t}} \bigg\}\bigg]. 
\end{align}

Define the following function,  
\begin{equation}
g_{\lambda_t}(\lambda_{t}) 
= \begin{cases}
(\lambda_t)^{-1} f(\widetilde{\sigma}^2_{t+1} |y_t, \lambda_{t+1}, \widetilde{\sigma}^2_{t}, \lambda_t, \bm{\vartheta}) f(\widetilde{\sigma}^2_{t} |\lambda_t, \bm{\vartheta})
& \text{ for } t=1, \\
(\lambda_t)^{-1} f(\widetilde{\sigma}^2_{t+1} |y_t, \lambda_{t+1}, \widetilde{\sigma}^2_{t}, \lambda_t, \bm{\vartheta}) \\
\hspace{20pt} \times f(\widetilde{\sigma}^2_{t} |y_{t-1}, \lambda_{t}, \widetilde{\sigma}^2_{t-1}, \lambda_{t-1}, \bm{\vartheta})
& \text{ for } t=2, \ldots, n-1, \\
(\lambda_t)^{-1} f(\widetilde{\sigma}^2_{t} |y_{t-1}, \lambda_{t}, \widetilde{\sigma}^2_{t-1}, \lambda_{t-1}, \bm{\vartheta}) 
& \text{ for } t=n, \end{cases}
\end{equation}
and note that I can approximate $g_{\lambda_t}$ to a function proportional to the probability density of log-normal distribution: 
\begin{align}
g_{\lambda_t}(\lambda_{t} ) 
&\approx (\lambda_t)^{-1} \exp \bigg\{ -\frac{1}{2s_{\lambda_t}}(\log\lambda_t -m_{\lambda_t})^2 \bigg\} \times\text{constant}, \\
s_{\lambda_t} &=\begin{cases}
(s_{\lambda_t, 1}^{-1} +\omega_{\eta\eta,0}^{-1})^{-1} & \text{ if } t=1, \\
(s_{\lambda_t, 1}^{-1} +\omega^{(\eta\eta)})^{-1} & \text{ if } t=2, \ldots, n-1, \\
\omega_{\eta\eta} -\omega_{\eta\epsilon}^2 & \text{ if } t=n, 
\end{cases} \\
m_{\lambda_t} &=\begin{cases}
s_{\lambda_t}(s_{\lambda_t, 1}^{-1}m_{\lambda_t,1}+\omega_{\eta\eta,0}^{-1}m_{\lambda_t,2}) & \text{ if } t=1, \\
s_{\lambda_t}(s_{\lambda_t, 1}^{-1}m_{\lambda_t,1}+\omega^{(\eta\eta)}m_{\lambda_t,2}) & \text{ if } t=2, \ldots, n-1, \\
m_{\lambda_t, 2} & \text{ if } t=n, 
\end{cases} \\
m_{\lambda_t, 2} &=\begin{cases}
 \log\widetilde{\sigma}^2_{t} & \text{ if } t=1, \\
 \log\widetilde{\sigma}^2_{t} -\phi (\log\widetilde{\sigma}^2_{t-1}-\log\lambda_{t-1}) -\omega_{\eta\epsilon} \frac{y_{t-1}\lambda_{t-1}^{\frac{1}{2}}}{\widetilde{\sigma}_{t-1}} & \text{ if } t=2, \ldots, n. 
 \end{cases}
\end{align}

I can show that a random variable with the gamma density function with parameters $(\alpha_{\lambda_t}, \beta_{\lambda_t})$, 
has mean $m_{\lambda_t}$ and variance $s_{\lambda_t}$, 
where $\alpha_{\lambda_t} =(\exp(s_{\lambda_t})-1)^{-1}, \;\; \beta_{\lambda_t} =\exp(-m_{\lambda_t}-\frac{1}{2}s_{\lambda_t}) (\exp(s_{\lambda_t})-1)^{-1}$. 
I observe that the log-normal density function is approximated by the gamma distribution with parameters $(\alpha_{\lambda_t}, \beta_{\lambda_t})$.

Then, $f(\lambda_{t} |\{y_s\}_{s=1}^n, \{r_s\}_{s=1}^n, \{\widetilde{\sigma}^2_{s}\}_{s=1}^n, \{\lambda_{s}\}_{s\neq t}, \bm{\vartheta})$ is approximated to the probability density function 
$f_a(\lambda_{t} |\{y_s\}_{s=1}^n, \{r_s\}_{s=1}^n, \{\widetilde{\sigma}^2_{s}\}_{s=1}^n, \{\lambda_{s}\}_{s\neq t}, \bm{\vartheta})$, where
\begin{align}
&f_a(\lambda_{t} |\{y_s\}_{s=1}^n, \{r_s\}_{s=1}^n, \{\widetilde{\sigma}^2_{s}\}_{s=1}^n, \{\lambda_{s}\}_{s\neq t}, \bm{\vartheta}) 
\propto 
\lambda_t^{\frac{\alpha_{t,1}}{2}-1}\exp\bigg( -\frac{\beta_{t, 1}}{2}\lambda_t \bigg), \\
&\alpha_{t,1}= \nu_1 +1 +2\alpha_{\lambda_t}, \;\; \beta_{t, 1} =\nu_2 + \frac{y_t^2}{\widetilde{\sigma}^2_t} +2\beta_{\lambda_t}. 
\end{align}

I generate a candidate $\lambda_t^\dagger \sim \mathrm{G}(\frac{\alpha_{t, 1}}{2}, \frac{\beta_{t, 1}}{2})$ and it is accepted with probability
\begin{equation}
\min\left\{ 1, \frac{g_{\lambda_t}(\lambda_{t}^\dagger ) \cdot \lambda_t^{\frac{\alpha_{\lambda_t}}{2}-1}\exp(-\frac{\beta_{\lambda_t}}{2}\lambda_t)}{g_{\lambda_t}(\lambda_{t} ) \cdot (\lambda_t^\dagger)^{\frac{\alpha_{\lambda_t}}{2}-1}\exp(-\frac{\beta_{\lambda_t}}{2}\lambda_t^\dagger)} \right\}, 
\end{equation}
where $\lambda_t$ is a current MCMC sample. 

\subsection{Generation of $\bm{\vartheta}$}
\noindent
{\it Generation of $\phi$.} 
The conditional posterior density function of $\phi$ is given by 
\begin{equation}
f(\phi |\{y_t\}_{t=1}^n, \{\sigma^2_t\}_{t=1}^n, \Omega) 
\propto g(\phi) \times\exp\bigg\{ -\frac{(\phi -m_\phi)^2}{2s_\phi} \bigg\}, 
\end{equation}
where
\begin{align}
&s_\phi = \Bigg\{\sum_{t=1}^{n-1} (\log\sigma^2_t)^2 \omega^{(\eta\eta)} \Bigg\}^{-1}, \;\;
m_\phi = s_\phi \Bigg\{\sum_{t=1}^{n-1} \omega^{(\eta\eta)}\bigg(\log\sigma^2_{t+1} -\omega_{\eta\epsilon}\frac{y_t}{\sigma_t}\bigg) \log\sigma^2_t \Bigg\}, \\
&g(\phi) = \pi(\phi) (\omega_{\eta\eta, 0})^{-\frac{1}{2}} \exp\bigg\{ -\frac{(\log\sigma^2_1)^2}{2\omega_{\eta\eta, 0}} \bigg\}. 
\end{align}
Sample $\phi^\dagger\sim\mathrm{TN}_{(-1,1)}(m_\phi, s_\phi)$ as a candidate, where a normal distribution with mean $m_\phi$ and variance $s_\phi$ truncated to the region $(-1, 1)$ is denoted as $\mathrm{TN}_{(-1,1)}(m_\phi, s_\phi)$. 
It is accepted with probability $\min\{1, g(\phi^\dagger)/g(\phi)\}$, where $\phi$ is a current MCMC sample.

\vspace{0.5\baselineskip}

\noindent
{\it Generation of $\Omega$.} 
I can obtain the conditional posterior density of $(\omega^{(\epsilon\eta)}, \omega^{(\eta\eta)})$ as follows:   
\begin{align}
&f(\omega^{(\epsilon\eta)}, \omega^{(\eta\eta)} |\{y_t\}_{t=1}^n, \{\sigma^2_t\}_{t=1}^n, \phi) \notag \\
&\propto \pi(\omega^{(\epsilon\eta)}, \omega^{(\eta\eta)}) \times f(\eta_0|\bm{\vartheta}) \prod_{t=1}^{n-1} f(y_t, \sigma^2_{t+1} |\sigma^2_{t}, \bm{\vartheta}) \\
&\propto (\omega_{\eta\eta, 0})^{-\frac{1}{2}} \exp\bigg\{ -\frac{(\log\sigma^2_1)^2}{2\omega_{\eta\eta, 0}} \bigg\} 
\times (\omega^{(\eta\eta)})^{\frac{n_1}{2}-1} \exp\bigg\{ -\frac{\omega^{(\eta\eta)}}{2s_1} \bigg\} \notag \\
&\hspace{20pt}\times (\omega^{(\eta\eta)})^{-\frac{1}{2}} \exp\bigg\{ -\frac{(\omega^{(\epsilon\eta)}-\omega^{(\eta\eta)}\delta_1)^2}{2\gamma_1\omega^{(\eta\eta)}} \bigg\}, 
\end{align}
where
\begin{align}
&n_1=n_0+n-1, \;\; s_1=(s_0^{-1} +\xi_{22} +\delta_0^2\gamma_0^{-1} -\delta_1^2\gamma_1^{-1})^{-1} \\
&\gamma_1 =(\gamma_0^{-1} +\xi_{11})^{-1}, \;\; \delta_1 =(-\xi_{21} +\delta_0\gamma_0^{-1})\gamma_1, \\
&\Xi =\begin{pmatrix} 
\xi_{11} & \xi_{12} \\
\xi_{21} & \xi_{22} 
\end{pmatrix}
=\sum_{t=1}^{n-1} 
\begin{pmatrix} 
\sigma^{-1}_t y_t \\ \eta_t 
\end{pmatrix}
\begin{pmatrix} 
\sigma^{-1}_t y_t \\ \eta_t 
\end{pmatrix}'.  
\end{align}
(For the multivariate case, see \cite{Ishihara_etal(2016)}.) 
$\Omega^\dagger$ is generated as a candidate as the following:

\begin{enumerate}
\item 
Generate $(\omega^{(\eta\eta)})^\dagger \sim\mathrm{G}\big(\frac{n_1}{2}, \frac{1}{2s_1}\big)$.
\item 
Generate $(\omega^{(\eta\epsilon)})^\dagger \sim\mathrm{N}(\omega^{(\eta\eta)})^\dagger \delta_1,  \gamma_1\omega^{(\eta\eta)})^\dagger)$.
\item 
Compute $\omega_{\eta\epsilon}^\dagger =-((\omega^{(\eta\eta)})^\dagger)^{-1} (\omega^{(\eta\epsilon)})^\dagger,$ 
$\omega_{\eta\eta}^\dagger = ((\omega^{(\eta\eta)})^\dagger)^{-1} + (\omega_{\eta\epsilon}^\dagger)^2$ 
and they are accepted with probability
\begin{equation}
\min \Bigg[ 1, 
(\omega_{\eta\eta, 0}^\dagger)^{-\frac{1}{2}} \exp\bigg\{ -\frac{(\log\sigma^2_1)^2}{2\omega_{\eta\eta, 0}^\dagger} \bigg\}
(\omega_{\eta\eta, 0})^{\frac{1}{2}} \exp\bigg\{ \frac{(\log\sigma^2_1)^2}{2\omega_{\eta\eta, 0}} \bigg\}
\Bigg]. 
\end{equation}
\end{enumerate}

\vspace{0.5\baselineskip}

\noindent
{\it Generation of $\bm{\nu}$.} 
For $\bm{\nu}$, the conditional posterior density is given by 
\begin{equation}
f(\bm{\nu} |\{\lambda_t\}_{t=1}^n)
\propto \pi(\bm{\nu}) \Bigg\{ \frac{(\frac{\nu_2}{2})^{\frac{\nu_1}{2}}}{\Gamma(\frac{\nu_1}{2})} \Bigg\}^n \prod \lambda_t^{\frac{\nu_1}{2}} \exp\Bigg( -\frac{\nu_2}{2} \sum_{t=1}^n \lambda_t \Bigg). 
\end{equation}
I transform $\bm{\nu}$ such that $\widetilde{\bm{\nu}}=(\log\nu_1, \log\nu_2)'$. 
Taylor expansion around the mode $\widehat{\bm{\nu}}$ gives the following approximation for the conditional posterior density of $\widetilde{\bm{\nu}}$ as 
\begin{equation}
f(\widetilde{\bm{\nu}} |\{\lambda_t\}_{t=1}^n)
\approx \text{constant} \times \exp\bigg\{ -\frac{1}{2}(\widetilde{\bm{\nu}} -\bm{m}_{\bm{\nu}})' S_{\bm{\nu}}^{-1} (\widetilde{\bm{\nu}} -\bm{m}_{\bm{\nu}}) \bigg\}, 
\end{equation}
where 
\begin{equation}
\bm{m}_{\bm{\nu}} =\widehat{\bm{\nu}} + S_{\bm{\nu}} \times\frac{\partial \log f(\widetilde{\bm{\nu}} |\{\lambda_t\}_{t=1}^n)}{\partial\widetilde{\bm{\nu}}}\Bigg|_{\widetilde{\bm{\nu}} =\widehat{\bm{\nu}}}, \;\;
S_{\bm{\nu}} =\Bigg[ -\frac{\partial^2 \log f(\widetilde{\bm{\nu}} |\{\lambda_t\}_{t=1}^n)}{\partial\widetilde{\bm{\nu}}\partial\widetilde{\bm{\nu}}'}\Bigg|_{\widetilde{\bm{\nu}} =\widehat{\bm{\nu}}} 
\Bigg]^{-1}.
\end{equation}
I generate a candidate $\widetilde{\bm{\nu}}^\dagger$ from the normal density $f_{\mathrm{N}}$ with mean vector $\bm{m}_{\bm{\nu}}$ and covariance matrix $S_{\bm{\nu}}$ and it is accepted with probability 
\begin{equation}
\min\bigg\{ 1, \frac{f(\widehat{\bm{\nu}}^\dagger |\{\lambda_t\}_{t=1}^n) f_{\mathrm{N}}(\widetilde{\bm{\nu}} |\bm{m}_{\bm{\nu}}, S_{\bm{\nu}})}{f(\widehat{\bm{\nu}} |\{\lambda_t\}_{t=1}^n) f_{\mathrm{N}}(\widetilde{\bm{\nu}}^\dagger |\bm{m}_{\bm{\nu}}, S_{\bm{\nu}})} \bigg\}. 
\end{equation}

\section{Empirical analysis}
\subsection{Data}
This section provides empirical results using asset returns and price ranges for two stock indices in the U.S. financial market: 
Dow Jones Industrial Average (DJIA) index and the S\&P500 index. 
The asset return $y_t$ is defined as $(\log p_{t} -\log p_{t-1})\times 100$, where $p_t$ is the asset's closing price at day $t$. 
The price range $r_t$ at day $t$ is calculated as $(\log H_t -\log L_t)\times 100$,  where $H_t$ is the highest and $L_t$ is the lowest asset prices at date $t$. 
For model comparison, I use corresponding five-minute sub-sampled realized volatility data obtained from Oxford-Man Institute's ``Realized Library'' (\cite{Heber_etal(2009)}).
The sample period is from January 3, 2012 to December 31, 2020 (2,254 observations for DJIA and 2,256 observations for S\&P500, respectively). 

\subsection{Estimation results}
I apply the return and price range dataset described in the last subsection to the proposed model with the following priors for the parameters:  
\begin{align}
&\frac{\phi +1}{2} \sim\mathrm{Be}(20, 1.5), \;\;
\omega^{(\eta\eta)}\sim\mathrm{G}\bigg(\frac{1}{2}, \frac{0.2}{2}\bigg), \;\; 
\omega^{(\epsilon\eta)}|\omega^{(\eta\eta)}\sim\mathrm{N}(0, 10\omega^{(\eta\eta)}), \\
&\nu_1 \sim\mathrm{G}\bigg( \frac{16}{2}, \frac{0.8}{2} \bigg), \;\;
\nu_2 \sim\mathrm{G}\bigg( \frac{16}{2}, \frac{0.8}{2} \bigg). 
\end{align}
Using the algorithm discussed in Section 3, 10,000 MCMC samples are generated after discarding the initial 1,000 samples as the burn-in. 

\begin{table} 
\caption{MCMC Estimation results. 
\label{table:4.2}} 
\begin{center}
\begin{tabular}{lrcr} \hline
Par. & Mean & 95\% interval 
& IF \\ \hline
\multicolumn{4}{l}{DJIA} \\
$\phi$ & 0.939 & (0.924, 0.954) 
& 15.7 \\
$\omega_{\epsilon\eta}$ & $-0.174$ & $(-0.201, -0.146)$ 
& 8.5 \\ 
$\omega_{\eta\eta}$ & 0.150 &  (0.120, 0.181) 
& 42.0 \\
$\nu_1$ & 17.663 & (13.757, 22.274) 
& 55.1 \\
$\nu_2$ & 27.549 & (20.742, 35.458) 
& 51.6 \\ \hline 
\multicolumn{4}{l}{S\&P500} \\
$\phi$ & 0.918 & (0.899, 0.935) 
& 13.9 \\
$\omega_{\epsilon\eta}$ & $-0.217$ & $(-0.248, -0.185)$ 
& 6.0 \\
$\omega_{\eta\eta}$ & 0.215 & (0.175, 0.261) 
& 29.8 \\ 
$\nu_1$ & 19.972 & (15.338, 26.331) 
& 58.0 \\
$\nu_2$ & 28.204 & (21.634, 37.378) 
& 58.1 \\ \hline
\end{tabular}
\end{center}
\end{table}

Table \ref{table:4.2} reports the posterior means, 95\% credible intervals
and estimated inefficiency factors (IF).  
The inefficiency factor is defined as $1+2\sum_{k=1}^\infty \rho(k)$, where $\rho(k)$ is the autocorrelation function for the parameter at lag $k$ (see \cite{Chib(2001)}). 

\begin{figure}
\begin{center}
\caption{DJIA. Illustration of $y_t$, $r_t$, posterior means of $\sigma_t$'s and $\lambda_t$'s, and 95\% credible interval upper and lower bounds (the two outmost gray lines) of $\lambda_t$'s. \label{fig:4.2-a}}
\includegraphics[height=9cm]{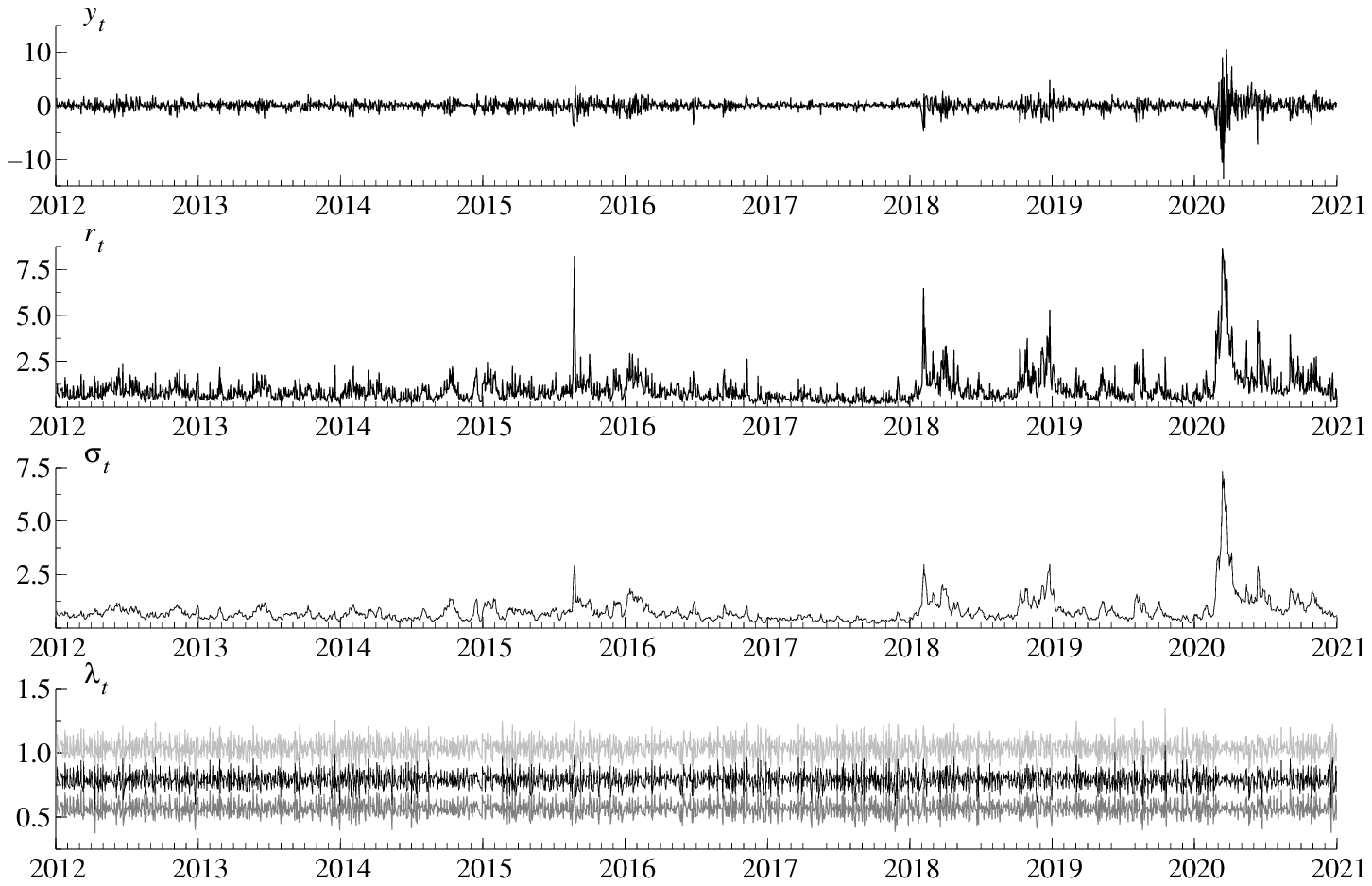}
\end{center}
\end{figure}

\begin{figure}
\begin{center}
\caption{S\&P500. Illustration of $y_t$, $r_t$, posterior means of $\sigma_t$'s and $\lambda_t$'s, and 95\% credible interval upper and lower bounds (the two outmost gray lines) of $\lambda_t$'s. \label{fig:4.2-b}}
\includegraphics[height=9cm]{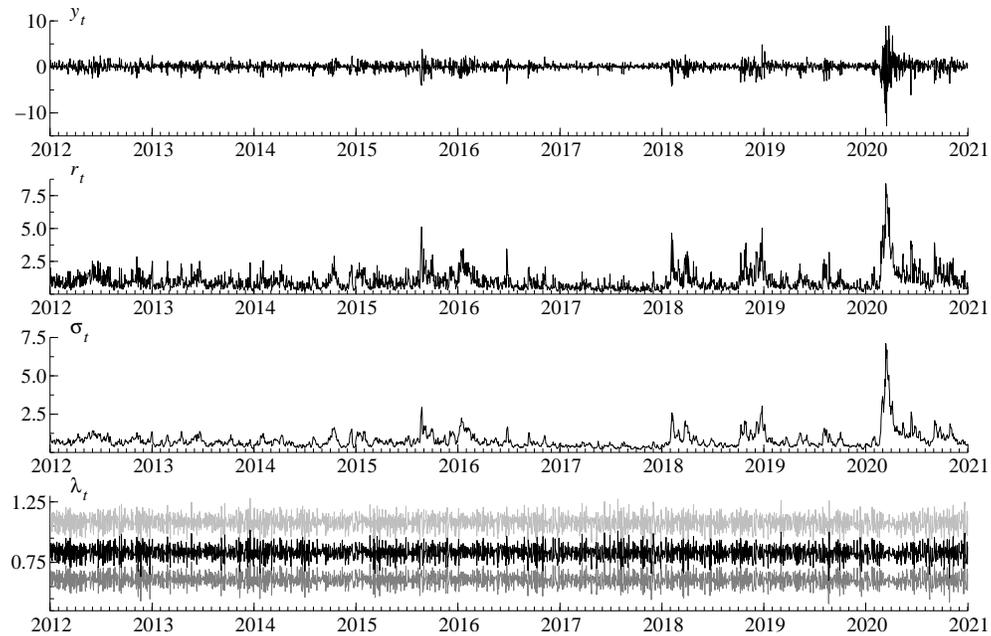}
\end{center}
\end{figure}

For DJIA (S\&P500) data, two top panels of Figure \ref{fig:4.2-a} (\ref{fig:4.2-b}) depicts the time series plot of $y_t$ and $r_t$, respectively. 
The transition for the estimated posterior means of $\sigma_t$ is illustrated in the third top panel of Figure \ref{fig:4.2-a} (\ref{fig:4.2-b}). 
The bottom panel of Figure \ref{fig:4.2-a} (\ref{fig:4.2-b}) reports the posterior means and the bounds for 95\% credible intervals of $\lambda_t$. 

The posterior means of the first order autoregressive parameter $\phi$ are over 0.9 and very high, which indicates that the processes of (log) volatilities are strongly persistent. 
It is consistent with the plot of the posterior means of $\sigma_t$'s recorded in Figures \ref{fig:4.2-a} and \ref{fig:4.2-b}. 
I observe the sharp increase of the trajectories of $\sigma_t$ in March 2020, resulted from COVID-19 pandemic. 
The trajectories increased in August 2015, corresponding to the global stock market sell-off while those increases in February 2018 and late 2018 are due to the spike in VIX and cryptocurrency crash, respectively. 

I notice that the 95\% credible intervals of $\omega_{\epsilon\eta}$ do not include $0$. 
For DJIA (S\&P500), the posterior mean of correlation coefficient between $y_t$ and $\log\sigma^2_{t+1}$ is equal to $-0.449$ ($-0.468$). 
It strongly suggests the existence of asymmetry or leverage effect and is consistent with reports in the vast amount of papers. 

When I use DJIA (S\&P500) data, the sample means of 95\% credible interval upper and lower bounds of $\lambda_t$'s are 1.037 (1.080) and 0.568 (0.606), respectively. 
I see that the 95\% credible intervals of the bias correction coefficients for the price range are large and that many of them include the value of 1. 
It implies that the measured price range contains a large amount of noise and that the downward bias for the range revealed in the previous studies is not necessarily significant. 

For DJIA (S\&P500), the acceptance rates of $\{\sigma^2_t\}_{t=1}^n$, $\{\lambda_t\}_{t=1}^{n}$, $\bm{\nu}$, $\phi$ and $\Psi$ in the independent chain Metropolis-Hastings algorithm are 0.952 (0.942), 0.957 (0.950), 0.985 (0.983), 0.984 (0.994) and 0.993 (0.993), respectively. 
I notice that the proposal density functions approximate the posterior accurately and that the MCMC algorithm can generate samples efficiently.

\subsection{Econometric model comparison}
\noindent
{\it Generation of volatility from posterior predictive distribution.} 
The posterior predictive density of $\log\sigma^2_{n+1}$ is given by
\begin{equation}
\log\sigma^2_{n+1} \sim\mathrm{N}(\phi\log\sigma^2_n +\omega_{\eta\epsilon}\sigma_{n}^{-1}y_n, \omega_{\eta\eta}-\omega_{\eta\epsilon}^2).
\end{equation}
I add a step of generation of $\sigma^2_{n+1}$ to each MCMC iteration described in the last section and obtain the estimated posterior predictive density. 

\vspace{0.5\baselineskip}

\noindent
{\it Model comparison based on robust loss function.} 
Suppose that $L(\sigma^2_t, h_{t})$ denotes a loss function, i.e., a measure for error between the true volatility of asset return at $t$-th day $\sigma^2_t$ and a forecast $h_{t}$. 
In addition, suppose that I have two volatility forecasts $h_{j,t}$ and $h_{k,t}$ based on two statistical models $j$ and $k$. 
$h_{j,t}$ outperforms $h_{k,t}$ with respect to $L$ if the expectation of the loss function is smaller than the other one, i.e., $\mathrm{E}\{L(\sigma^2_t, h_{j,t})\} < \mathrm{E}\{L(\sigma^2_t, h_{k,t})\}$. 
Needless to say, the true volatility is not observed in the financial market, and I have to use a volatility proxy. 
Since a volatility proxy contains noise, it is not always true that the ranking between two volatility forecasts $h_{j,t}$ and $h_{k,t}$ is preserved if one substitute a conditionally unbiased proxy $\widehat{\sigma}^2_t$ to the true volatility, i.e.,  
\begin{equation}
\mathrm{E}\{L(\widehat{\sigma}^2_t, h_{j,t})\} < \mathrm{E}\{L(\widehat{\sigma}^2_t, h_{k,t})\} \Longleftrightarrow
\mathrm{E}\{L(\sigma^2_t, h_{j,t})\} < \mathrm{E}\{L(\sigma^2_t, h_{k,t})\}.
\end{equation}

By \cite{Patton(2011)}, it was shown that a class of loss functions holds equivalence of such ordering between two volatility forecasts.  
The following two loss functions mean squared error (MSE) and quasi-likelihood (QLIKE) are included in \cite{Patton(2011)}'s class: 
\begin{align}
&\mathrm{MSE}: L(\widehat{\sigma}^2_t, h_t) =\frac{(h_t -\widehat{\sigma}^2_t)^2}{2},  \\
&\mathrm{QLIKE}: L(\widehat{\sigma}^2_t, h_t) =\frac{\widehat{\sigma}^2_t}{h_t} -\log\frac{\widehat{\sigma}^2_t}{h_t} -1.
\end{align}
Note that these take value of zero if and only if $h_t =\widehat{\sigma}^2_t$.  
\cite{Patton(2011)} used realized measure and Parkinson's range-based estimator $r_t^2/(4\log 2)$ as conditionally unbiased volatility proxies. 

It is well known that realized measure and range-based estimator for variance of asset return often take smaller value than the true due to the market microstructure noise or overnight effect as stated in Section 1. 
\cite{HansenLunde(2006)} recommended that realized volatility $RV_t$ should be scaled using daily returns as 
\begin{equation}
RV_t^{\mathrm{scale}} =c_{\mathrm{scale}}RV_t, \;\; c_{\mathrm{scale}}=\frac{\sum_{s=1}^n (y_t -n^{-1}\sum_{s'=1}^ny_{s'})^2}{\sum_{s=1}^nRV_t}.
\end{equation}
It implies that the mean of scaled realized volatility $RV_t^{\mathrm{scale}}$ takes the same value of the sample variance of corresponding daily returns. 

In this subsection, I use realized volatility (RV) and Parkinson's range-based estimator (RG) as volatility proxies which are scaled as \cite{HansenLunde(2006)}. 
Posterior predictive mean of $\sigma^2_{t+1}$ is adopted as the forecasts. 
Thus, I may obtain the ranking between two volatility forecasts, which is consistent with that using true volatility with respect to the loss functions described above. 

Although the ranking provided as above is robust to the noise in the volatility proxy and is reliable, 
one may notice that the difference between the two losses is not measured. 
I use a predictive ability test proposed by \cite{GiacominiWhite(2006)} to determine whether the difference is statistically significant. 
I utilize constant and lagged loss difference for the test function and conduct the test in line with the theorem in that publication.  

\vspace{0.5\baselineskip}

\noindent
{\it Empirical results.} 
The rolling forecast is conducted as follows: 
(i) Apply the estimation algorithm to the data from 2012 to 2018 to (I set the data as $\{y_t\}_{t=1}^n$ and $\{r_t\}_{t=1}^n$) and generate the parameters and unobserved variables as well as one-step-ahead forecast of variance, $\sigma^2_{n+1}$. 
(6,000 MCMC samples are generated after discarding 1,000 samples from the iterations.)
(ii) The new data observed at the followed day is added to the period and the data observed at the oldest day is removed. 
(iii) Generate the model parameters and unobserved variables for the data (relabeled as $\{y_t\}_{t=1}^n$ and $\{r_t\}_{t=1}^n$). Further, generate $\sigma^2_{n+1}$. 
(iv) Go to Step (ii). 
In the end, I obtain posterior predictive means of $\sigma^2_{n+1}$ (499 in total for DJIA data and 497 in total for S\&P500 data, respectively). 

As a benchmark model, I employ RSV model (\cite{Takahashi_etal(2009)}). 
\cite{Takahashi_etal(2021)} conducted model comparison between RSV model and the other competing volatility models and show that it has an excellent predictive ability. 
Details for this benchmark model are stated in Appendix. 
I use daily asset return data and associated realized measure of variance as mentioned in Section 5.1 for this competing model. 

\begin{table} [H]
\caption{DJIA. Average loss and $p$-values of Giacomini and White's test with the null hypothesis that the predictive ability is equal to that of SVRG model. \label{table:empirical2}} 
\begin{center}
\begin{tabular}{crrrrrrrr} \hline
\multicolumn{1}{c}{} & \multicolumn{4}{c}{RV} & \multicolumn{4}{c}{RG} \\ 
\multicolumn{1}{c}{} & \multicolumn{2}{c}{MSE} & \multicolumn{2}{c}{QLIKE} & \multicolumn{2}{c}{MSE} & \multicolumn{2}{c}{QLIKE} \\ 
Model & Ave. loss & $p$-value & Ave. loss & $p$-value & Ave. loss & $p$-value & Ave. loss & $p$-value  \\ \hline
SVRG & 8.873 & & 0.224 & & 6.376 & & 0.423 & \\
RSV & 13.107 & 0.062 & 0.239 & 0.341 & 10.001 & 0.125 & 0.461 & 0.296 \\
\hline
\end{tabular}
\end{center}
\end{table}

\begin{table} [H]
\caption{S\&P500. Average loss and $p$-values of Giacomini and White's test with the null hypothesis that the predictive ability is equal to that of SVRG model. \label{table:empirical3}} 
\begin{center}
\begin{tabular}{crrrrrrrr} \hline
\multicolumn{1}{c}{} & \multicolumn{4}{c}{RV} & \multicolumn{4}{c}{RG} \\ 
\multicolumn{1}{c}{} & \multicolumn{2}{c}{MSE} & \multicolumn{2}{c}{QLIKE} & \multicolumn{2}{c}{MSE} & \multicolumn{2}{c}{QLIKE} \\ 
Model & Ave. loss & $p$-value & Ave. loss & $p$-value & Ave. loss & $p$-value & Ave. loss & $p$-value  \\ \hline
SVRG & 12.461 & & 0.272 & & 4.566 & & 0.426 & \\
RSV & 16.149 & 0.109 & 0.274 & 0.340 & 5.794 & 0.306 & 0.423 & 0.314 \\
\hline
\end{tabular}
\end{center}
\end{table}

Tables \ref{table:empirical2} and \ref{table:empirical3} report the average values of the MSE and QLIKE loss functions of the forecasts. 
I use RV and RG as the volatility proxies. 
Tables \ref{table:empirical2} and \ref{table:empirical3} also show the $p$-values of Giacomini and White's test with the null hypothesis that the predictive ability is equal to that of SVRG model. 

For DJIA data, both MSE and QLIKE are lower for SVRG model than for RSV model. 
It indicates that SVRG model ranks better than RSV model, whereas the results of Giacomini and White's test show that significant difference is not found between the two models at significant level of 0.01. 
For S\&P500 data, Table \ref{table:empirical3} reports similar empirical results as DJIA data. 
In summary, I conclude that the predictive ability of the model is equivalent to or better than that of RSV model. 
Without high frequency data in the financial market, I may obtain an accurate forecast of volatility of an asset return by using the proposed model.

\newpage
\section{Conclusion}
This paper introduces a novel SV model in the state space form by adding new observation equation which relates the variance and the price range. 
I investigate the probability distribution of range for daily stock exchange and the rigorous sampling algorithm from the density function. 
An MCMC algorithm is developed based on the Bayesian framework to obtain estimation results for the proposed SVRG model. 

An empirical study is presented using daily returns and price ranges of DJIA and S\&P500 indices in the U.S. stock market. 
I find the persistence in the (log) volatility processes, the existence of the leverage effect, and a non-negligible amount of noise in the observed price range as suggested by the previous empirical papers. 
The model comparison is based on out-of-sample predictive performance 
and the results reveal that the forecasts generated by the model are as accurate (if not more so) as those of RSV model.

Although the proposed univariate model can be extended to a multivariate model, building such models presents various challenges. 
Many financial market researchers have found dynamic correlations between asset returns, which should be considered in the multivariate volatility model.  
I must guarantee the covariance matrix of multivariate asset returns is positive definite even if dynamic correlation structure is incorporated. 
Furthermore, there are usually too many model parameters and latent variables to obtain stable estimation results for such multivariate or high-dimensional time-varying covariance structure models.  
In the context of SV models, e.g., \cite{KuroseOmori(2016)} proposed parsimonious modeling of multivariate asset returns and it would be worthy of considering such extension. 
This is beyond the scope of this paper and the author leaves the extension to the future study.

\section*{Acknowledgement}
The author wishes to thank Kaoru Irie, Tsunehiro Ishihara, Yasuhiro Omori, and Makoto Takahashi for the comments. 
This research was supported by JSPS KAKENHI Grant Numbers 26245028, 19H00588, 20K19751. 
The computational results in this article were generated using Ox (see \cite{Doornik(2009)}). 

\section*{Appendix}
\subsection*{A.1 Realized stochastic volatility model}
RSV model is given in the state space form as
\begin{align}
y_t &= \exp\Big( \frac{\alpha_t}{2}  \Big) \epsilon_t, \;\; t=1, \ldots, n, \\
x_t &= \xi +\alpha_t + w_t,  \;\; t=1, \ldots, n, \\
\alpha_{t+1} &= (1-\phi)\mu +\phi\alpha_t +\eta_t,  \;\; t=1, \ldots, n-1, \\
\alpha_{1} &= \mu +\eta_1, \\
\begin{pmatrix} \epsilon_t \\ w_t \\ \eta_t \end{pmatrix} 
&\sim \mathrm{N} \left( 
\begin{pmatrix} 0 \\ 0 \\ 0 \end{pmatrix},  
\begin{pmatrix} 1 & 0 & \omega_{\epsilon\eta} \\
0 & \omega_{ww} & 0 \\
\omega_{\eta\epsilon} & 0 & \omega_{\eta\eta} \end{pmatrix} 
\right), \;\;
\eta_1 \sim\mathrm{N}(0, \omega_{\eta\eta}/(1-\phi^2)), 
\end{align}
where $y_t$ is the return, $\alpha_t$ the logarithm of variance and $x_t$ the logarithm of a realized measure of the variance. 

Note that the posterior predictive density of $\alpha_{n+1}$ is given by $\mathrm{N}((1-\phi)\mu+\phi\alpha_n +\omega_{\eta\epsilon} \exp(-\frac{1}{2}\alpha_n)y_n)$.  

\cite{KuroseOmori(2020)} extended this model to the multivariate case and developed an efficient Bayesian estimation method. 
I assume that the prior distributions of the parameters as \cite{KuroseOmori(2020)} and employ the estimation algorithm for the univariate RSV model. 

\bibliographystyle{chicago}
\bibliography{de_a}
\end{document}